\documentclass[prl,showpacs,superscriptaddress,amsfonts,twocolumn]{revtex4}
\usepackage{bm}
\usepackage{times}
\usepackage[dvips]{graphicx}

\begin{document}

\title{Bailout Embeddings, Targeting of KAM Orbits,
and the Control of Hamiltonian Chaos}

\author{Julyan H. E. Cartwright} 
\homepage{http://lec.ugr.es/~julyan}
\email{julyan@lec.ugr.es}
\affiliation{Laboratorio de Estudios Cristalogr\'aficos, CSIC, 
E-18071 Granada, Spain}
\author{Marcelo O. Magnasco} 
\homepage{http://asterion.rockefeller.edu/}
\email{marcelo@sur.rockefeller.edu}
\affiliation{Mathematical Physics Lab, Rockefeller University, 
Box 212, 1230 York Avenue, NY 10021}
\author{Oreste Piro}
\homepage{http://www.imedea.uib.es/~piro}
\email{piro@imedea.uib.es} 
\affiliation{Institut Mediterrani d'Estudis Avan\c{c}ats, CSIC--UIB,
E-07071 Palma de Mallorca, Spain}

\date{version of \today}
 
\begin{abstract}
We present a novel technique, which we term bailout embedding, that can be 
used to target orbits having particular properties out of all orbits in a flow
or map. We explicitly construct a bailout embedding for Hamiltonian systems so
as to target KAM orbits. We show how the bailout dynamics is able to lock onto
extremely small KAM islands in an ergodic sea.
\end{abstract}


\maketitle

Control of chaos in nonlinear dynamical systems has been achieved by applying
small perturbations that effectively change the dynamics of the system
around the region --- typically a periodic orbit --- that one wishes to 
stabilize \cite{OGY}. This method has been successful in dissipative systems,
but its extensions to control and targeting in 
Hamiltonian systems \cite{lai,schroer,wu,oloumi,macau,bolotin,zhang,zhang2,xuwang} 
have met various difficulties not
present in the dissipative case: the absence of attracting sets, for
instance, makes it hard to stabilize anything. In addition, these
methods require that one know in advance what one wants to do; in 
particular, the orbit to be stabilized may have to be known in
advance to a fair accuracy. 

In this Letter we present a novel technique allowing us to control 
Hamiltonian chaos, in such a way as to keep the original dynamics intact, 
but which shifts the stability of different kinds of orbits in the dynamics. 
We do so by embedding our Hamiltonian system within a larger space,
meaning we {\em augment} the number of degrees of freedom, keeping 
an intact copy of the original system on one privileged slice; all
of the control is achieved through use of the perpendicular directions
to this intact copy. We call this
method a bailout embedding for reasons that will promptly become clear.
We apply the technique to an extremely hard and hitherto close to intractable
problem in chaos control of Hamiltonian systems: selecting small KAM
(Kolmogorov--Arnold--Moser) islands within chaotic seas in systems that are
almost ergodic. We show below how our technique is able to find and render
asymptotically stable minute islands of order within a map. While formerly this
could be done by sophisticated algorithms on the basis of complex logic, our
method distills the complexity of this calculation into a simple,
forward-iterates dynamical system. Thus our method allows us to
stabilize KAM islands without knowing their location in advance.

\begin{figure*}[htbp]
\begin{center}
\includegraphics*[width=0.24\textwidth]{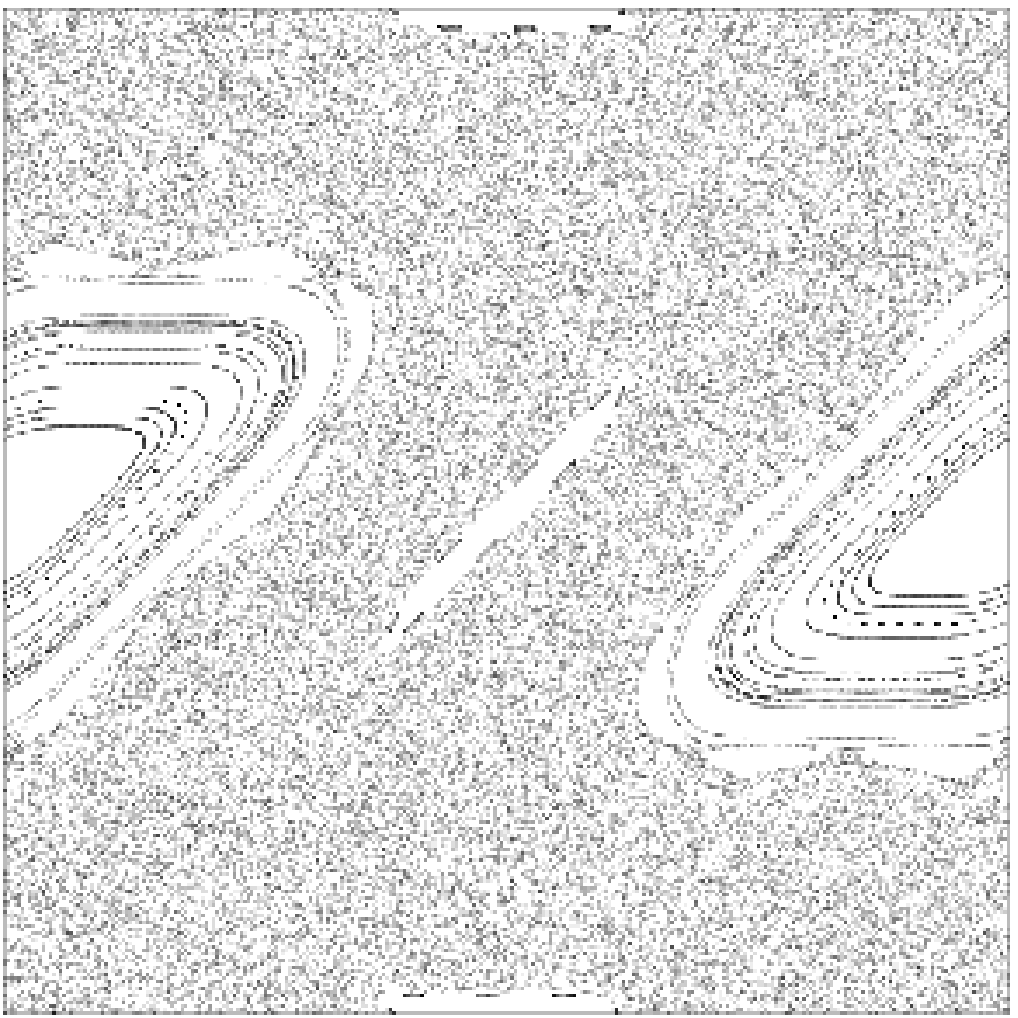}
\includegraphics*[width=0.24\textwidth]{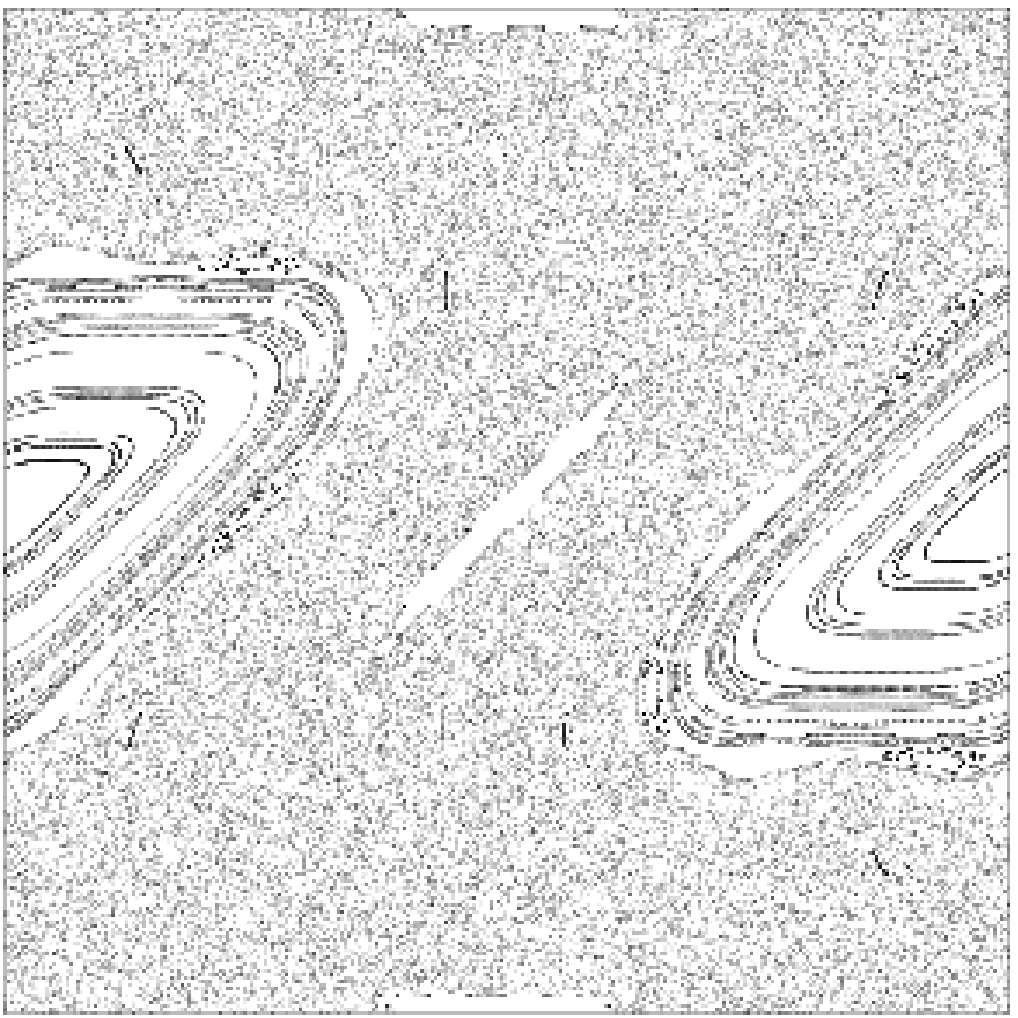}
\includegraphics*[width=0.24\textwidth]{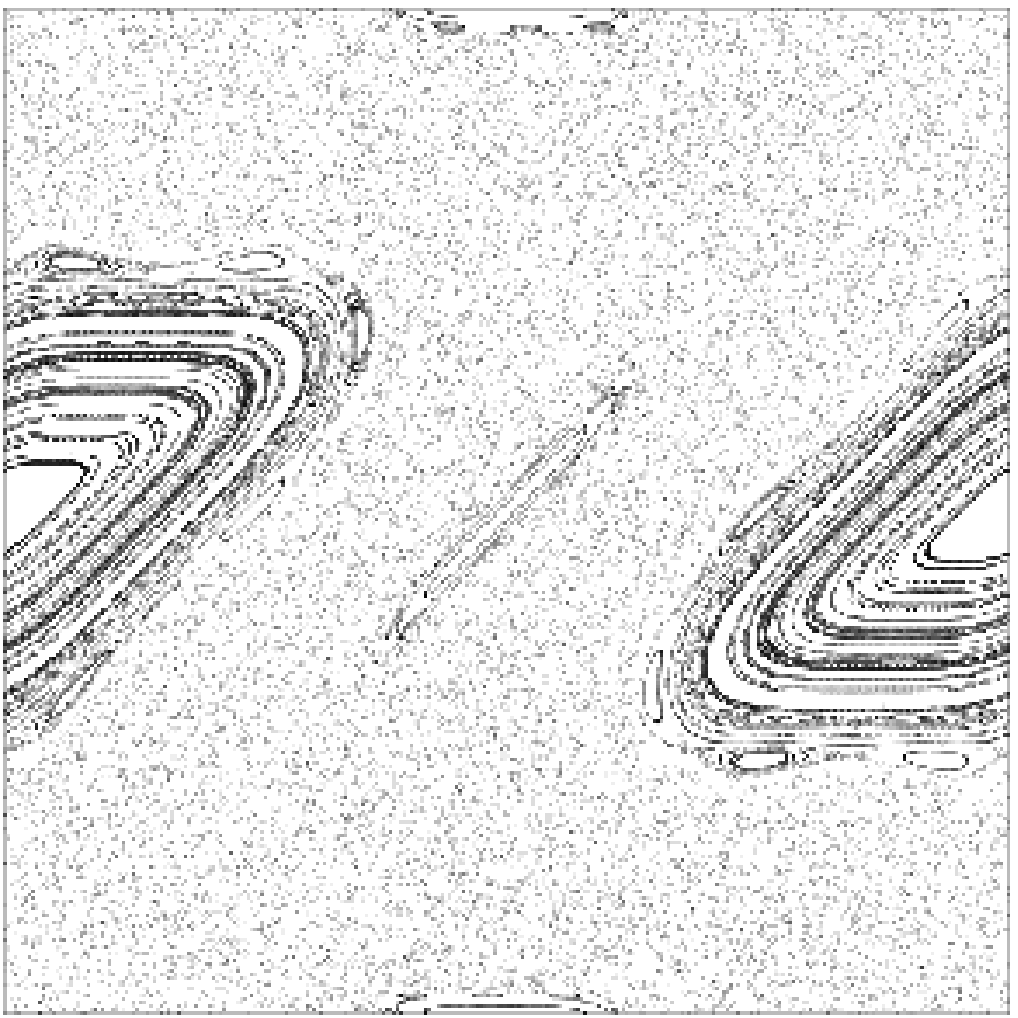}
\includegraphics*[width=0.24\textwidth]{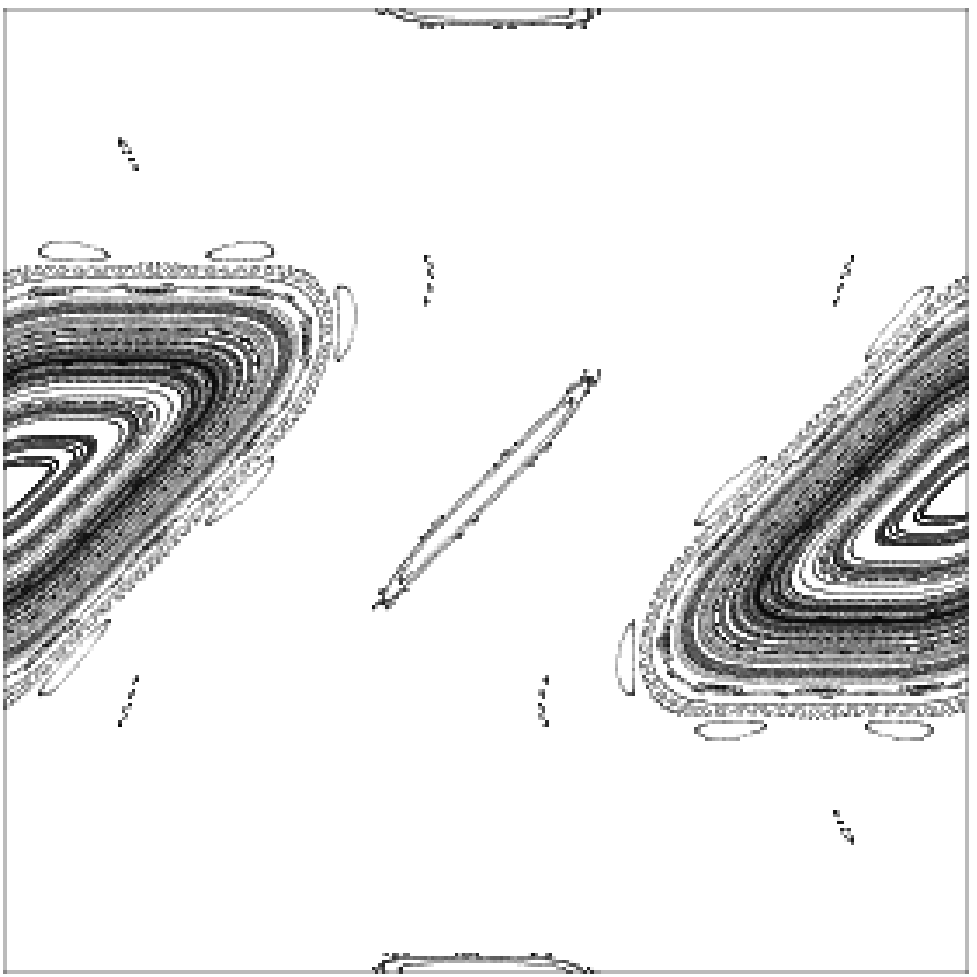}
\end{center}
\caption{\label{baildraw}
We illustrate how the bailout embedding can target KAM orbits using
the standard map at $k=2$.
100 random initial conditions were chosen, iterated for 10000 steps,
then the next 1000 iterations are shown. The squares represent the 
unit torus. 
(a) Original map, (b) $\lambda=1$, (c) $\lambda=0.6$, (d) $\lambda=0.5$ }
\end{figure*}

\begin{figure}[hbp]
\begin{center}
\includegraphics*[width=0.95\columnwidth]{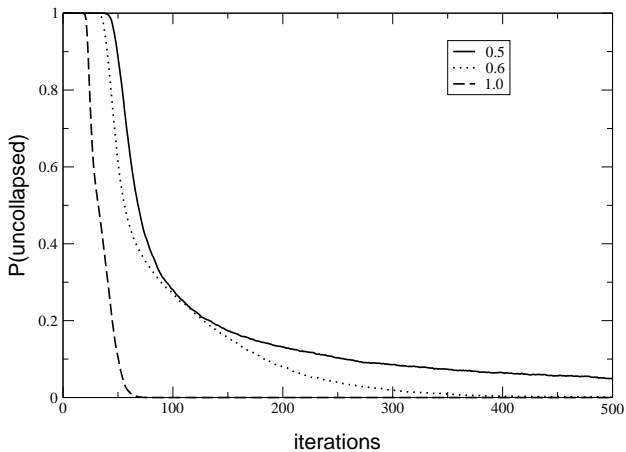}
\end{center}
\caption{\label{pirofig}
As time passes, random initial conditions eventually collapse onto 
the original dynamics. Here we plot the fraction of initial conditions
such that $|x_{n+1}-f(x_n)|>10^{-10}$ as a function of the number of
iterations elapsed, for 8000 random initial conditions, at $k=2$. The
three curves correspond to cases (bcd) in Figure 1.  
}
\end{figure}
 
A striking application of Eq.~(\ref{odebailout}) is to a case rather close to
the original setting of Eq.~(\ref{passive}): an incompressible fluid flow  $\bm
u$ translates in dynamical-systems terms to a divergence-free flow, of which
Hamiltonian systems are the most important class. A classical problem in
Hamiltonian dynamics is locating KAM tori. Hamiltonian systems live between two
opposite extremes, of fully integrable systems and fully ergodic ones. Fully
integrable systems are characterized by dynamics unfolding on invariant tori.
The KAM theorem asserts that as a parameter taking the system away from
integrability is increased, these tori break and give rise to chaotic regions
in a precise sequence; for any particular value of this parameter in a
neighborhood of the integrable case, there are surviving tori. The problem with
finding them is that, the dynamics being volume-preserving, merely evolving
trajectories either forwards or backwards does not give us convergence onto
tori \cite{gutzwiller},
and since for large values of the nonlinearity, they
cover a very small measure of the phase space, locating them becomes
an extremely difficult problem. Moreover, there is the further problem that 
even if we start on an island, we should be able to recognize it as such.
In fact, several sophisticated analytical and semianalytical techniques
had to be developed to assist in the search and characterization of KAM tori
in highly nonlinear Hamiltonian systems (see for instance \cite{percival3}). 
We shall now show that the bailout embedding solves these
problems by transforming the KAM trajectories into global attractors of the
embedded system; finding them is now independent of the choice of
initial conditions.

Our technique is based upon the dynamics of a small neutrally buoyant sphere
--- a passive scalar --- in an incompressible fluid flow $\bm u$
\cite{neutpartprl}. Under assumptions allowing us to retain only the Bernoulli,
Stokes drag, and Taylor added mass contributions to the force exerted by the
fluid on the sphere, the equation of motion for the sphere at the position $\bm
x$ is
\begin{equation}
\label{passive}
\frac{d}{dt}\left( \dot{\bm x}-{\bm u}({\bm x})\right) =
-(\lambda +{\bm\nabla\bm u})\cdot\left(\dot{\bm x}-{\bm u}({\bm x})\right) 
\end{equation}
so that the difference between the particle velocity and the velocity of the
surrounding fluid is exponentially damped with negative damping constant
$-(\lambda+{\bm\nabla\bm u})$. In the case in which the flow gradients
reach the magnitude of the viscous drag coefficient, there is the possibility
that around hyperbolic stagnation points the Jacobian matrix 
$\bm\nabla\bm u$ may acquire a positive
eigenvalue in excess of the drag coefficient. In these instances, the 
trajectories of these passive scalars, instead of converging exponentially 
onto 
$\dot{\bm x}={\bm u}$,
detach from such trajectories. The result is that the passive scalars
explore practically all of the flow, but tend to stay away from regions 
of high shear.

Equation \ref{passive} is but one instance of a more general structure. Let us
consider a differential equation of the form $\dot{x}=f(x)$. If we take its
time derivative, we obtain a different differential equation, 
$\ddot{x}=f'(x)\dot{x}$. It is different because, being second order, it exists
in a larger space and has many solutions that are not solutions of the smaller
one. Still, the original equation is contained within the larger system, in the
sense that every solution of $\dot{x}=f(x)$ is a solution of 
$\ddot{x}=f'(x)\dot{x}$. We may say that $\dot{x}=f(x)$ is embedded within
$\ddot{x}=f'(x)\dot{x}$. There are infinitely many ways to embed; for example, 
$\ddot{x}=f'(x)f(x)$ is also an embedding, but it is clearly inequivalent to
the first.

Of course, embedding a system changes notions of stability, because stability
refers to perturbations, and in a larger system there are all of the old
perturbations plus a batch of new ones. 
So, even though all of the solutions of the original system are preserved,
by adding new directions away from the old solutions we may transform formerly
stable solutions into unstable ones in the larger setting. See, for example,
studies of the manifold bubbling transition \cite{venkataramani}. 
The trivial way to embed a system is through a cross product; for instance,
$\dot{x}=f(x),\, x\in\mathbb{M}$ is embedded within 
$\mathbb{M}\times\mathbb{R}$ as
\begin{eqnarray}
\dot{x} & = & f(x)+g(x,y), \nonumber \\
\dot{y} & = & \alpha y
,\end{eqnarray}
where $g(x,y)$ is arbitrary except for requiring that 
$g(x,0)\equiv 0$, which guarantees that for 
$y=0$ we have the original system.
If $\alpha<0$ then $y$
always dies out, and so we always recover the original object; in this
case, we can call the embedding itself stable, in the sense that any motion 
away from the embedded object takes us back.
Both of the ``derivative'' embeddings in the previous paragraph were unstable;
stable versions can be constructed rather simply, for 
instance 
\begin{equation}
\frac{d}{dt}\left( \dot{x}-f(x)\right) =-k\left( \dot{x}-f(x)\right)
,\label{embed}\end{equation}
of which the previous examples were the 
$k=0$ limit.
This embedding ensures that the distance between the actual trajectory
and the embedding diminishes exponentially with time for any initial condition.
Equation (\ref{embed}) begins to resemble Eq.~(\ref{passive}).

We define a {\em bailout embedding} as one of the form
\begin{equation}
\label{odebailout}
\frac{d}{dt}\left( \dot{x}-f(x)\right) =-k(x)\left( \dot{x}-f(x)\right) 
,\end{equation}
where $k(x)<0$ on a set of orbits that are unwanted, and $k(x)>0$ otherwise.
Thus the natural behavior of a bailout embedding is that the trajectories in
the full system tend to detach or bail out from the embedded subsystem
into the larger
space, where they bounce around. If, after bouncing around for a while, these
orbits reach a stable region of the embedding, $k(x)>0$, they will once again
collapse onto the original dynamical system. In this
way we can create a larger version of the dynamics in which specific sets of
orbits are removed from the asymptotic set, while preserving the dynamics of
another set of orbits --- the wanted or targeted one --- as attractors of the 
enlarged
dynamical system. For the special choice of $k(x)=-(\lambda +\nabla f)$, as in
the fluid dynamics of a passive scalar described by Eq.~(\ref{passive}),  these
dynamics were shown in \cite{neutpartprl} to detach from saddle points and
other unstable regions in conservative dynamics.

\begin{figure*}[htbp]
\begin{center}
\includegraphics*[width=0.24\textwidth]{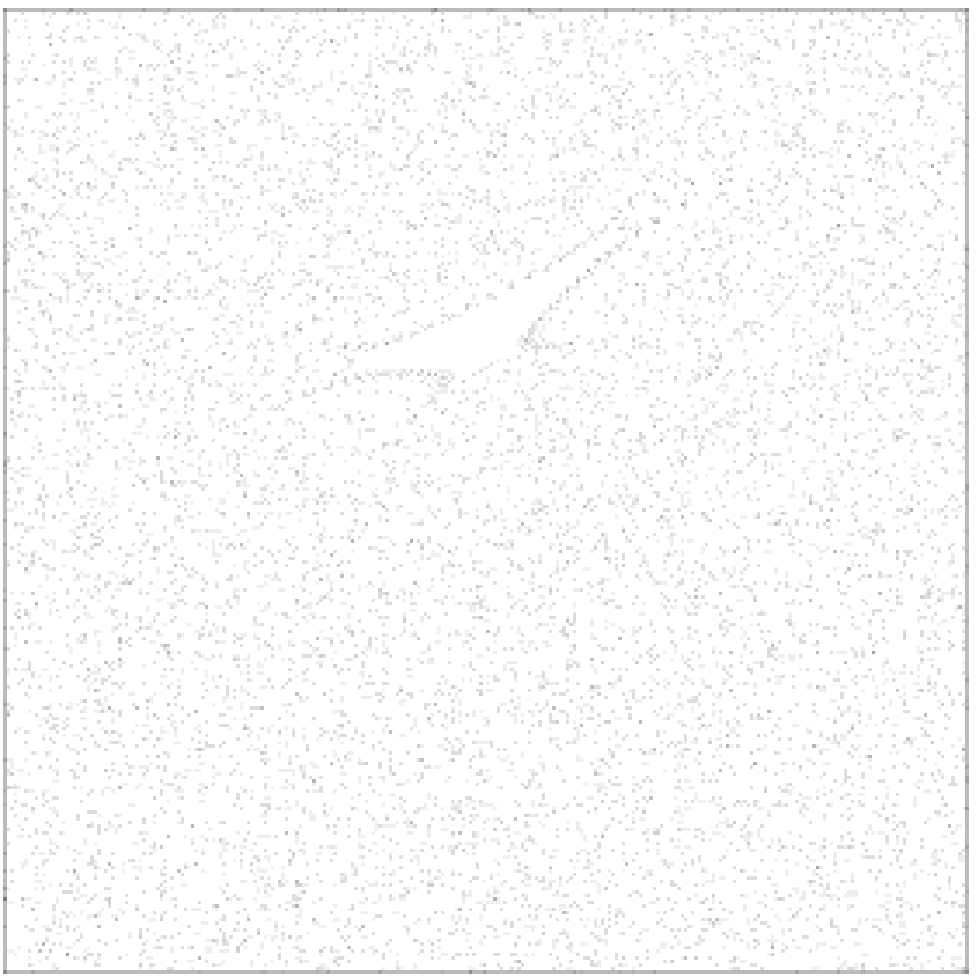}
\includegraphics*[width=0.24\textwidth]{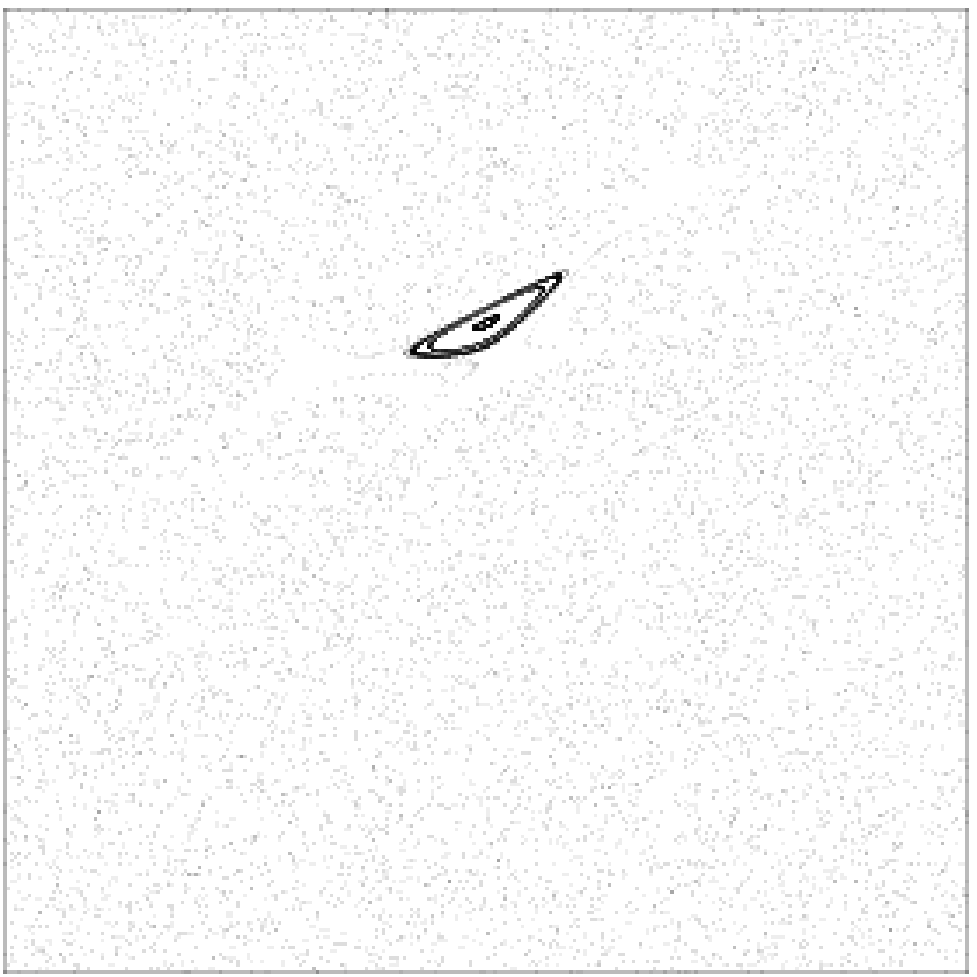}
\includegraphics*[width=0.24\textwidth]{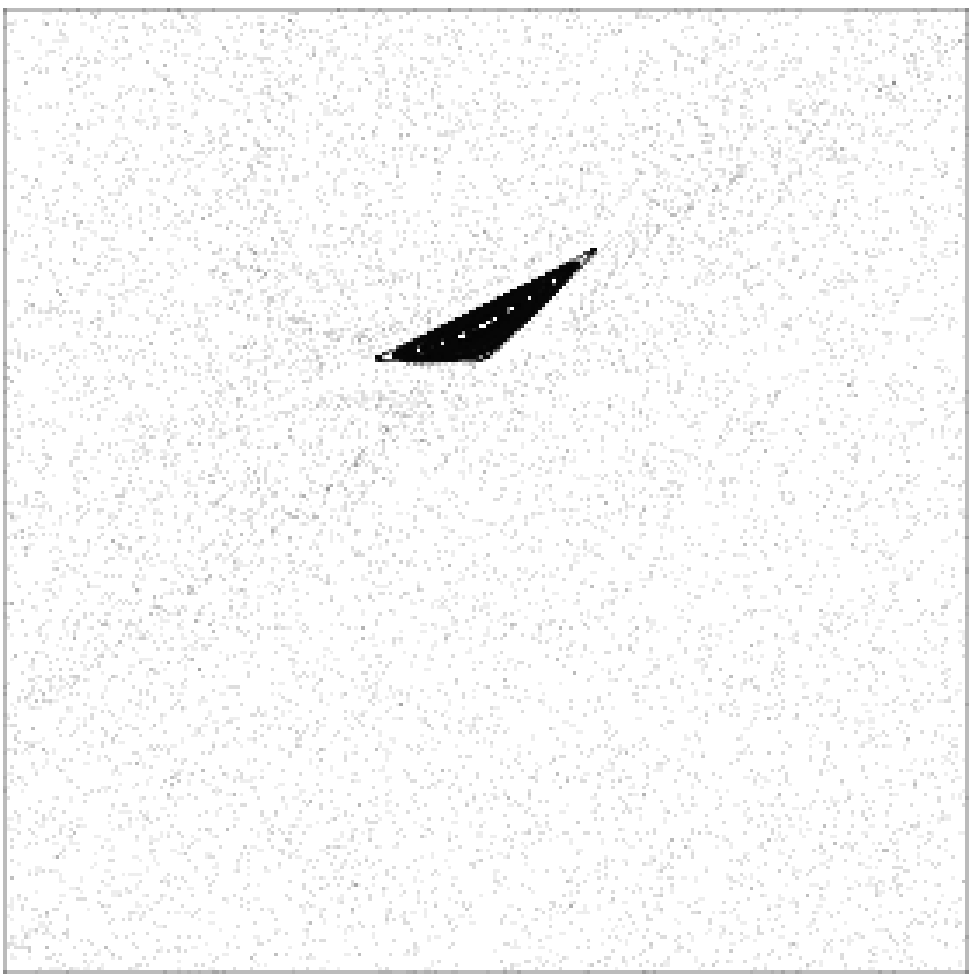}
\includegraphics*[width=0.24\textwidth]{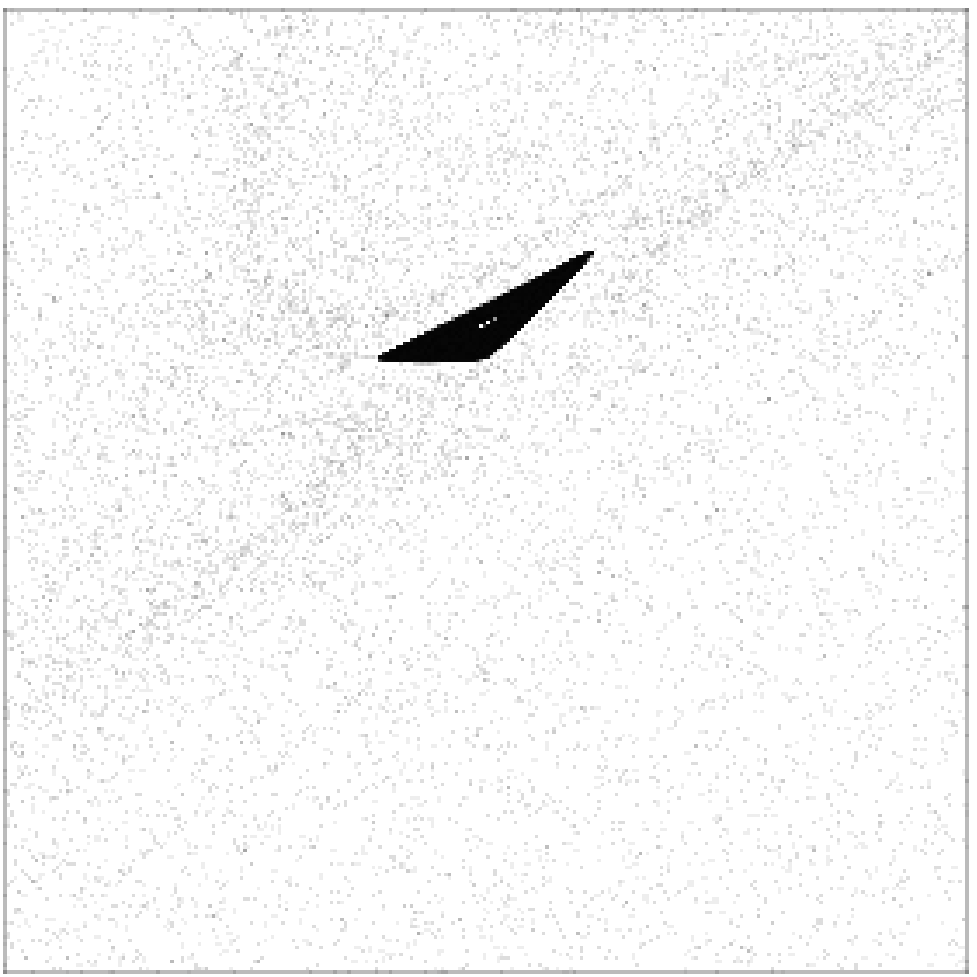}
\end{center}
\caption{\label{seven}
The bailout process can find small KAM islands. The standard map for $k=7$ 
has a chaotic sea covering almost the entire torus, except for a tiny 
period 2 KAM island near position $0,2/3$. As in Figure 1, except:
the squares here are a box $-0.05<x<0.05$, $0.61<p<0.71$, 
1000 random initial conditions and 20000 stabilization iterations. 
(a) Original map, (b) $\lambda=1.4$, (c) $\lambda=1.3$, (d) $\lambda=1.2$. 
}
\end{figure*}

It is not hard to extend flow bailout,
Eq.~(\ref{odebailout}), to maps in the obvious fashion. Given a map 
$x_{n+1}=T(x_{n})$ the bailout embedding is given by 
\begin{equation}
x_{n+2}-T(x_{n+1})=K(x_{n})(x_{n+1}-T(x_{n}))
\label{map_bailout}\end{equation}
provided that $|K(x)|>1$ over the unwanted set. (In the map system, almost any
expression written for the ordinary differential equation translates to
something close to an exponential; in particular, stability eigenvalues have to
be negative in the ordinary differential equation case to represent stability,
while they have to be smaller than 1 in absolute value in the map case). The
particular choice of the gradient as the bailout function $k(x)=-(\lambda
+\nabla f)$ in a flow translates in the map setting to  $K(x)=e^{-\lambda
}\nabla T$. A classical testbed of Hamiltonian systems is the standard map, an
area-preserving map introduced by Taylor and Chirikov. The standard map is
given by 
\begin{eqnarray}
x_{n+1} & = & x_{n}+\frac{k}{2\pi}\sin (2\pi y_{n}) \nonumber \\
y_{n+1} & = & y_{n}+x_{n+1}
\label{std_map}\end{eqnarray}
where $k$ is the parameter controlling integrability.

In order to embed the standard map, we only need to replace $T$ and $K(x)$ in 
Eq.~(\ref{map_bailout}) 
with the appropriate expressions. $T$ stems directly from Eq.~(\ref{std_map})
and, in
accordance with the previous definitions, $K(x)$ becomes:
\begin{equation}
K(x) = \left(
\begin{array}{cc}
1 & k \cos 2 \pi y_n \\
1 & k \cos 2 \pi y_n + 1 
\end{array}
\right)
\end{equation}
Notice that due to the area preserving property of the standard map, the
two eigenvalues of the derivative matrix must multiply to one. If they
are complex, this means that both have an absolute value of one, while if
they are real, generically one of them will be larger than one and the
other smaller. We can then separate the phase space into elliptic and
hyperbolic regions corresponding to each of these two cases.
If a trajectory of the original map lies
entirely on the elliptic regions, the overall factor $\exp(-\lambda)$ will
damp any small perturbation away from it in the embedded system. But for
chaotic trajectories, which inevitably must visit some hyperbolic regions,
there exists a value of $\lambda$ such that perturbations away from a standard
map-trajectory are amplified instead of dying out in the embedding.
As a consequence, trajectories are effectively expelled from the chaotic
regions to finally settle in the safely elliptic KAM islands. This process
can be observed clearly in Figure 1. As the value of $\lambda$ is 
decreased, the number of trajectories starting from random initial 
conditions which eventually settle into the 
KAM tori increases; see Figure 2. 

This process is sensitive enough to work even at hitherto intractably
high nonlinearities. For $k=7$, the standard map is {\sl almost} ergodic: 
it covers almost the entire torus with a single chaotic orbit. Only a 
tiny island of irreductible order resists this invasion. It is located 
around $(0,0.6774)$ and covers an area approximately $2\times 10^{-5}$ 
of the torus. 
Thus, from random initial conditions one would expect to see this island
only once every 50000 attempts. Figure 3 shows how easily our bailout 
method finds this island from simple forward iterates. 

A Hamiltonian system does not usually just satisfy volume conservation, 
but also will conserve the Hamiltonian itself. Given an ordinary differential 
equation $\dot x = f(x)$ with a conserved quantity $E\equiv 0$,
then $f\cdot \nabla E=0$. However, building
a bailout embedding by the procedure above does not lead to dynamics that 
satisfy $\dot E\equiv 0$, because the bailout embedding should be $2n-2$ 
dimensional.
This is clearly undesirable in the case of Hamiltonian systems, so we
show now how to derive a bailout embedding that will obey a conservation law. 
The bailout equation can be written 
$$  \ddot x = (\nabla f -\lambda )\cdot  (\dot x - f) + \nabla f \cdot \dot x$$
We need to correct this acceleration so that it stays on the second tangent
space of the $E\equiv 0$ surface. Let us call the raw bailout 
acceleration $u$. 
The second derivative $\ddot x$ has to satisfy
$ \ddot x \cdot \nabla E + \dot x \cdot \nabla \nabla E \cdot \dot x = 0 $, so 
we can modify $u$ to
$$ \ddot x = u - { u \cdot \nabla E \over | \nabla E | ^2 } \nabla E 
+ { \nabla E \over | \nabla E | }  \dot x \cdot \nabla \nabla E \cdot \dot x
$$
This equation, given that we start on $\dot x \cdot \nabla E=0$, will 
then preserve this property. 

We have presented a novel method in the control and targeting of chaos in 
nonlinear 
dynamical systems, the bailout embedding. While potentially useful in 
any dynamical-system setting, this method is especially suited
to Hamiltonian systems. Unlike other chaos-control strategies,
this method does not obliterate the original dynamics of the system, but
rather preserves it in a privileged slice of phase space embedded in 
a higher-dimensional space, and merely 
shifts around the stability of its orbits. A suitable choice of a bailout
function allows this strategy to target a complex set of orbits. We 
have shown, in particular, the targeting of KAM orbits, a case well-known
from classical studies to be especially hard. 

We should like to thank Leo Kadanoff and Marcelo Viana for useful discussions. 
JHEC acknowledges the financial support of the Spanish CSIC, Plan Nacional del
Espacio contract ESP98-1347.
MOM acknowledges the support of the Meyer Foundation.
OP acknowledges the Spanish Ministerio de Ciencia
y Tecnologia, Proyecto CONOCE, contract BFM2000-1108.

\end{document}